\documentclass[10pt,twocolumn]{article} 
\usepackage{arxiv/simpleConference}
\usepackage{times}
\usepackage{graphicx}
\usepackage{amssymb}
\usepackage{url,hyperref}

\usepackage{cite}
\usepackage{hyperref}
\usepackage{amsmath,amssymb,amsfonts}
\usepackage{algorithmic}
\usepackage{graphicx}
\usepackage{xcolor}
\usepackage{xspace}
\usepackage{listings}
\usepackage{balance}
\usepackage{todonotes}
\usepackage{multirow}
\usepackage{tcolorbox}

\usepackage[textfont={small,it}]{caption}
\definecolor{codegreen}{rgb}{0,0.6,0}
\definecolor{codegray}{rgb}{0.5,0.5,0.5}
\definecolor{codepurple}{rgb}{0.58,0,0.82}
\definecolor{backcolour}{rgb}{0.95,0.95,0.92}

\lstdefinestyle{mystyle}{
    backgroundcolor=\color{backcolour},   
    commentstyle=\color{codegreen},
    keywordstyle=\color{magenta},
    numberstyle=\tiny\color{codegray},
    stringstyle=\color{codepurple},
    basicstyle=\ttfamily\footnotesize,
    breakatwhitespace=false,         
    breaklines=true,                 
    captionpos=b,                    
    keepspaces=true,                 
    numbersep=5pt,                  
    showspaces=false,                
    showstringspaces=false,
    showtabs=false,                  
    tabsize=2
}

\usepackage{csquotes}
\usepackage{cleveref}
\usepackage{balance}
\usepackage{tabularx}

\usepackage{dblfloatfix}

\usepackage[numbers]{natbib}

\def\BibTeX{{\rm B\kern-.05em{\sc i\kern-.025em b}\kern-.08em
    T\kern-.1667em\lower.7ex\hbox{E}\kern-.125emX}}
    
\begin{document}

\title{On the use of test smells for prediction of flaky tests}

\author{
    \textbf{Bruno Henrique Pachulski Camara} \\
    \textit{Department of Computer Science} \\
    \textit{Federal University of Paraná} \\
    Curitiba, PR, Brazil \\
    \texttt{bhpachulski@ufpr.br} \\

\and

    \textbf{Marco Aurélio Graciotto Silva}\\
    \textit{Department of Computing} \\
    \textit{Federal University of Technology - Paraná}\\
    Campo Mourão, PR, Brazil \\
    \texttt{magsilva@utfpr.edu.br} \\
    
\and

    \textbf{Andre Takeshi Endo} \\
    \textit{Department of Computing} \\
    \textit{Federal University of Technology - Paraná} \\
    Cornélio Procópio, PR, Brazil \\
    \texttt{andreendo@utfpr.edu.br} \\

\and

    \textbf{Silvia Regina Vergilio} \\
    \textit{Department of Computer Science} \\
    \textit{Federal University of Paraná} \\
    Curitiba, PR, Brazil \\
    \texttt{silvia@inf.ufpr.br} \\

}

\maketitle

\begin{abstract}

Regression testing is an important phase to deliver software with quality. However, flaky tests hamper the evaluation of test results and can increase costs.
This is because a flaky test may pass or fail non-deterministically and to identify properly the flakiness of a test requires rerunning the test suite multiple times.
To cope with this challenge, approaches have been proposed based on prediction models and machine learning. 
Existing approaches based on the use of the test case vocabulary may be context-sensitive and prone to overfitting, presenting low performance when executed in a cross-project scenario.  
To overcome these limitations, we investigate the use of test smells as predictors of flaky tests.  
We conducted an empirical study to understand if test smells have good performance as a classifier to predict the flakiness in the cross-project context, and analyzed the information gain of each test smell. 
We also compared the test smell-based approach with the vocabulary-based one. As a result, we obtained a classifier that had a reasonable performance (Random Forest, 0.83) to predict the flakiness in the testing phase. 
This classifier presented better performance than vocabulary-based model for  cross-project prediction. 
The {\it Assertion Roulette} and {\it Sleepy Test} test smell types are the ones associated with the best information gain values.

\noindent \textbf{Kewords:} test flakiness, regression testing, test smells, machine learning

\end{abstract}

\maketitle

\section{Introduction}
\label{sec:intro}

Regression testing is an important phase to deliver software continuously with quality and minimizing failures after changes in the production code. In this phase, developers evaluate the test results to decide whether the program has a bug as a consequence of recent changes. However, the existence of flaky tests makes this evaluation untrustworthy.  This happens because some tests have an intermittent behavior, that is, they  pass and fail when executed in the same codebase~\cite{Luo-etal:2014, HerzigNagappan2015}.  This non-deterministic behavior, alternating between passing and failing without any code changes, frustrates developers. Flaky tests are challenging to debug, and a single failing test can halt release cycles~\cite{Alshammari-etal:2021}. 

Studies from literature show that flaky tests are common and appear in most large-scale projects~\cite{Eck-etal:2019}. 
In such cases, developers may spend important resources in analyzing failures that are due to flaky tests and not to actual problems in the code.  Practitioners get now used to rerun each newly observed failure several times, to ascertain that it is a genuine regression failure and not an intermittent one~\cite{Pinto-etal:2020}. This negatively impacts the productivity and contributes to increase costs.

A recent review on flaky tests points a growing interest in this subject ~\cite{Zolfaghari-etal:2020}. The academic and industrial software development communities have worked towards the definition of flaky tests~\cite{Luo-etal:2014, Micco:2016, Palmer:2019, Lam-etal:2019, WingLam-etal:2020}, reporting their occurrence~\cite{Eck-etal:2019, WingLam-etal:2019:issta}, building datasets~\cite{Bell-etal:2018, Lam-etal:2019, Pinto-etal:2020, Alshammari-etal:2021}, and finding automatically ways to identify them~\cite{TariqKing-etal:2018, Moran-etal:2020, Ahmad2020, Pinto-etal:2020, Camara-etal:2021, Haben2021, Alshammari-etal:2021, Haben2021}. The identification of flaky tests is usually addressed by two kinds of approaches. Dynamic approaches usually re-execute the test cases a fixed number of times~\cite{Lam-etal:2019, Shi-etal:2019}. This is expensive and error-prone. Moreover, it is not easy to determine how many executions is enough. This can be challenging  and an inadequate choice can lead to false negatives.  Static approaches in turn do not require code re-execution~\cite{Luo-etal:2014, Eck-etal:2019, Shi-etal:2019, Pinto-etal:2020, Atif-etal:2017, TariqKing-etal:2018}.

Most of the works apply Machine Learning (ML) methods to build models to predict flakiness likelihood~\cite{Alshammari-etal:2021,Atif-etal:2017,Camara-etal:2021,Pinto-etal:2020,TariqKing-etal:2018}.  These works differ on the adopted ML method and features used as predictors. Some of them use as predictors features obtained through the execution of code, such as coverage and runtime~\cite{Alshammari-etal:2021,TariqKing-etal:2018}.  

Models built using only static features present many advantages and are less costly~\cite{Pinto-etal:2020}. 
Considering this advantage, Pinto et al.~\cite{Pinto-etal:2020} build the set of predictors considering that there are some patterns within the test code that may be employed to automatically identify flaky tests. Then, tokens are identified and processed by using Natural Language Processing (NLP) techniques to synthesize a vocabulary of flaky tests. This vocabulary is used as predictors in combination with some common static test case features regarding number of lines of code and occurrence of certain Java keywords. This work applied five ML algorithms and Random Forest and Support Vector Machine (SVM) reached the best performance (F-measure of 0.95). 

The works of \citet{Ahmad2020} and \citet{Haben2021} attempted to reproduce the findings  of Pinto et al. and evaluated the effectiveness of vocabulary-based features in Python projects. 
\citet{Haben2021} also used an information retrieval technique to statically estimate testing coverage, and investigated  whether it could improve the performance of a vocabulary-based model, but without success. 
\citet{Camara-etal:2021} replicated the work of Pinto et al. and also investigated the generalization of the original results in real scenarios. By using different datasets, the authors evaluated the performance of the vocabulary-based model for a cross-project context, considering  intra- and inter-project test flakiness prediction. The authors concluded that the vocabulary-based approach is context-sensitive and prone to overfitting, presenting low performance when executed in a cross-project scenario. 

Considering these negative results, this work investigates the use of an alternative approach for flaky test prediction, based on test smells.  Similarly to the concept of bad code smells~\cite{Fowler:2020}, test smells are associated to potential design problems in the test code, that is, problems related to the way test cases are organized, implemented and interact with each other~\cite{Deursen-etal:2001}. Test smells may also impact the software quality and has been recently associated to test flakiness. 
For instance, \citet{Alshammari-etal:2021} investigated the use of test smells as predictors of flaky tests but in combination with other features. The set of features used included static metrics such as number of lines and assertions but also dynamics metrics related to line coverage, what can increase costs. 

Unlike the work of  Alshammari et al. we adopted a set of predictors only composed by metrics collected statically. In addition to the size of test case and number of smells in the test code, we also adopted binary features related to the presence or not of 19 test smells. Moreover, we investigated the use of the obtained models for cross-project prediction in comparison with the vocabulary-based approach. To this end, we employed the same dataset and algorithms used in our previous work~\cite{Camara-etal:2021}. As a result, we obtained a classifier that has a reasonable performance (Random Forest, 0.83\%).
This classifier presented better performance than vocabulary-based model for  cross-project prediction. The {\it Assertion Roulette} and {\it Sleepy Test} smells are the ones associated with the best information gain values. In this way, the main contributions of this paper are:

\begin{itemize}
    \item An empirical study of the adoption of test smell-based models for the prediction of test flakiness. Such an approach presents some advantages. The use of smells can be collected statically and automatically. In our work we use the tsDetect tool~\cite{Peruma-etal:2020}; 
    
    \vspace{0.2cm}
    \item Results about the importance of the different type of smells and their relation with flaky tests. This can guide testers to better design test code;
    
    \vspace{0.2cm}
    \item Discussion on some implications and limitations about the generalization of the results and obtained models for different projects in a cross-validation scenario, that can guide future research in the area; and
    
    \vspace{0.2cm}
    \item A repository containing the procedures, datasets, and scripts generated from this study  at \url{https://github.com/bhpachulski/SAST21-Paper}.
\end{itemize}

The paper is organized as follows. Section~\ref{sec:background} provides background about test smells and flaky tests. Section~\ref{sec:motivating} contains an example showing how smells and flakiness may relate. This example  serves as motivation to our work. Section~\ref{sec:relatedWork} reviews related work. Section~\ref{sec:methodology} describes the methodology adopted in our study. Section~\ref{sec:results} presents and analyses the obtained results. Section~\ref{sec:threats} discusses the main threats of our study. Section~\ref{sec:conclusion} presents our final remarks and concludes the paper.

\section{Background}
\label{sec:background}

In this section we provide an overview of the main topics related to this work: flaky tests and test smells. 

\subsection{Flaky test}

In regression testing, test suites are executed to validate if code changes, like new features added or bug fixing, do not  negatively impact the software.  However, not all test failures during the regression testing uncover new faults on production code~\cite{HerzigNagappan2015}. Some tests have an intermittent behavior, that is,  in a given moment they execute with success, but in others they fail for the same code version. A test with this characteristic, which passes/fails non-deterministically, is known as flaky~\cite{Luo-etal:2014}.

Figure~\ref{javacode-example} illustrates a flaky test reported by \citet{Lam-etal:2019}. A time limit (\texttt{{\small timeout=2000}}) to execute the method \texttt{{\small testIssue}} is defined in Line~1.  In this way, if the method takes more than two seconds (i.e., 2000 ms), the test case fails. In some cases, the test case is executed with success, but eventually, if the resource that is required is not ready, the test case will fail. 
In other words, the test case behavior can be associated with the environment, hardware, and some other reasons~\cite{Lam-etal:2019}. The existence of flaky test is inconvenient for software development and regression testing activities. 
Many times, to reproduce the failure that is non-deterministic, and to detect  flakiness it is necessary to re-execute the code and this can be costly for the software development process.

\begin{lstlisting}[language=Java,  label={javacode-example}, caption=Flaky test example~\cite{Lam-etal:2019}.]
@Test(timeout=2000)
public void testIssue() throws Exception {
  final int port = SocketUtil.getAvailablePort(); 
  WebSocketServer server = new WebSocketServer( 
    new InetSocketAddress(port)) { ...}
  ...
}
\end{lstlisting}

\citet{Luo-etal:2014} identified a list of the most prominent categories of flaky tests for developers and researchers to focus on. As a  result of this study  a classification of the main causes of flakiness was introduced. Figure~\ref{tab:causes-flakiness} describes the causes found. 
The most prevalent root causes are \textit{Async Wait} with 45\% of the cases, \textit{Concurrency} with 20\%, and \textit{Test Order Dependency} with 12\%. These three causes represent 77\%  of the cases~\cite{Luo-etal:2014}. 

\begin{table*}[h]
\centering
\caption{Classification for the causes of flakiness, proposed by~\citet{Luo-etal:2014}.}
\label{tab:causes-flakiness}
\begin{tabularx}{\textwidth}{|l|X|}

\hline
\textbf{Cause}             & \textbf{Description} \\ \hline

{\it Async Wait}               & When the test makes an asynchronous call and does not wait for the result to be available before using it. \\ \hline
{\it Concurrency }              & When the test starts several threads that interact non-deterministically, causing a undesirable behavior.        \\ \hline
Test Order Dependency     & When the test outcome depends on the order in which the test cases are run.                                  \\ \hline
{\it Resource Leak }            & When the test does not acquire or release resources, e.g., memory allocations or database connections.  \\ \hline
{\it Network }                  & When the test execution depends on the network and can be flaky because the network is hard to control and unpredictable.     \\ \hline
{\it Time }                     & When the test assertion is based on a time or due to the precision by which time is reported as it can vary inter-platforms. \\ \hline
{\it IO operations}             & When the test needs to access an IO resource that may also cause flakiness.                              \\ \hline
{\it Randomness}                & When the test assertion is based on a random number or information, this may be a potential flaky.                \\ \hline
{\it Floating Point Operations} & When dealing with floating point operations is known to lead to tricky non-deterministic cases.         \\ \hline
{\it Unordered Collections}     & When iterating over unordered collections the code should not assume that the elements are returned in a particular order. \\ \hline

\end{tabularx}%
\end{table*}

\subsection{Test smells}

Test code, just like any production code, is subject to be poorly written, without taking good programming practices into account, fact that introduces the so-called anti-patterns or smells~\cite{Fowler:2020}. 
Test smells are a deviation of how the tests  should  be written, organized and how tests  should interact with others. That deviation can indicate test design problems, and can hurt the test performance~\cite{Deursen-etal:2001, Peruma-etal:2019}. This section gives an overview of bad code smells that are specific for test code. 

Initially, \citet{Deursen-etal:2001} proposed a set of test smells composed of {\it Assertion Roulette}, {\it Eager Test}, {\it General Fixture}, {\it Lazy Test}, {\it   Mystery Guest}, {\it Resource Optimism}, and {\it Sensitive Equality}. Then, \citet{Peruma-etal:2019} extended the types of test smells with others inspired by bad test programming practices mentioned in unit testing based literature. In a later work, Peruma~et~al.t~\cite{Peruma-etal:2020} introduced tsDetect, an open source test smells detection tool. \Cref{tab:test-smells} presents definitions and detection rules of test smells used by tsDetect.  
This state-of-the-art tool detects a comprehensive set of test smells and was adopted in our study.

\begin{table*}[h]
\centering
\caption{Test smells and detection rules of tsDetect~\cite{Peruma-etal:2020}.}
\label{tab:test-smells}
\begin{tabularx}{\textwidth}{|l|X|}

\hline
\multicolumn{1}{|l}{\textbf{Test Smell}} & \multicolumn{1}{|l|}{\textbf{Detection Rule}} \\ \hline

{\it Assertion Roulette} & It occurs when a method has more than an assertion, so if one fails it is difficult to define which one.  \\ \hline
{\it Conditional Test Logic} & A test method that contains control flow statements (i.e \textsc{if}, \textsc{switch}, conditional expression, \textsc{for}, \textsc{foreach} and \textsc{while} statements).  \\  \hline
{\it Constructor Initialization} & A test class that contains a constructor declaration.  \\ 
{\it Default Test} & A test class is named either ``ExampleUnitTest'' or ``ExampleInstrumentedTest''.  \\  \hline
{\it Duplicate Assert} & A test method that contains more than one assertion statement with the same parameters.  \\  \hline
{\it Eager Test} & A test method that contains multiple calls to multiple production methods.  \\  \hline
{\it Empty Test} & A test method that does not contain a single executable statement.  \\  \hline
{\it General Fixture} & Not all fields instantiated within the \textsc{setUp} method of a test class are utilized by all test methods in the same test class.  \\  \hline
{\it Ignored Test} & A test method or class that contains the @Ignore annotation.  \\  \hline
{\it Lazy Test} & Multiple test methods calling the same production method.  \\  \hline
{\it Magic Number Test} & An assertion method that contains a numeric literal as an argument.  \\  \hline
{\it Mystery Guest} & A test method containing object instances of files and databases classes.  \\  \hline
{\it Redundant Print} & A test method that invokes either the print or \textsc{println} or \textsc{printf} or write method of the System class.  \\  \hline
{\it Redundant Assertion} & A test method that contains an assertion statement in which the expected and actual parameters are the same.  \\  \hline
{\it Resource Optimism} & A test method that utilizes an instance of a File class without calling the \textsc{exists()}, \textsc{isFile()} or \textsc{notExists()} methods of the object.  \\  \hline
{\it Sensitive Equality} & A test method that invokes the \textsc{toString()} method of an object.  \\  \hline
{\it Sleepy Test} & A test method that invokes the \textsc{Thread.sleep()} method.  \\  \hline
{\it Unknown Test} & A test method that does not contain a single assertion statement and @Test(expected) annotation parameter.  \\  \hline
{\it Verbose Test} & A test that is too long and hard to understand~\cite{Meszaros:2007, Spadini-etal:2020}. \\ \hline

\end{tabularx}
\end{table*}

\section{Motivating Example}
\label{sec:motivating}

Considering the causes of flakiness described by~\citet{Luo-etal:2014} and test smells, we can observe it is possible to make relations between some of them. 
For example, instructions that inject delays may be related to the flakiness root causes \textit{Async Wait} or \textit{Concurrency}; it is possible to wait for an {\textsc async} operation to complete or provide delays to synchronize two or more threads. 
The use of delays is related to the \textit{Sleepy Test} smell.

To illustrate this, Figure~\ref{javacode-example-sleep} shows a method from the test class \textsc{TestMemoryLocks}, from the open-source project Oozie\footnote{\url{https://oozie.apache.org/}}, a workflow scheduler system to manage Apache Hadoop\footnote{\url{https://hadoop.apache.org/}} jobs. This example is a flaky test, extracted from the dataset provided by \citet{Pinto-etal:2020}.  
Using the tsDetect tool~\cite{Peruma-etal:2020} in the test method \textsc{testReadWriteLock}, two test smells are detected: \textit{Sensitive Equality}, and \textit{Sleepy Test}.

\begin{lstlisting}[language=Java, label={javacode-example-sleep}, caption=A flaky test from the Oozie project~\cite{Pinto-etal:2020}.]
@Test
public void testReadWriteLock() throws Exception {
    StringBuffer sb = new StringBuffer("");
    Locker l1 = new ReadLocker("a", 1, -1, sb);
    Locker l2 = new WriteLocker("a", 2, -1, sb);
    
    new Thread(l1).start();
    Thread.sleep(500);
    new Thread(l2).start();
    Thread.sleep(500);
    l1.finish();
    Thread.sleep(500);
    l2.finish();
    Thread.sleep(500);
    assertEquals("a:1-L a:1-U a:2-L a:2-U", sb.toString().trim());
}
\end{lstlisting}

The test starts instantiating a \textsc{StringBuffer} object to be used by the code under test and on the assertion (Line~3). Then, two objects from classes \textsc{ReadLocker} and \textsc{WriteLocker} are instantiated (Lines~4 and 5); these objects consist of the main production code under test. 
A thread with \textsc{ReadLocker} is started in Line~7 and, after a sleep (Line~8). The same occurs to the \textsc{WriteLocker} object  (\mbox{Lines~9 and 10}). 
Similarly, both locker objects are finished (Lines~11 to 14). Finally, the assertion checks whether the simulated behavior with read and write operations to variable \texttt{sb} is accurate or not (Line~15). 

The root cause for flakiness is probably related to \textit{Concurrency} due to the use of delays associated with the threads. 
This test code contains the smell \textit{Sleepy Test}, since it uses the \textsc{sleep} method on a \textsc{thread}. 
Also, the \textit{Sensitive Equality} smell is identified by the use of a \textsc{toString} method in the assertion. 
As defined by \citet{Luo-etal:2014}, the root cause of flakiness \textit{Concurrency} happens when the test starts several threads that interact non-deterministically, causing the intermittent behavior. 

Just like this example, other four test cases are identified as flaky in the same class; their structure are pretty similar.
This example indicates that the root cause of a flaky test may be associated to the presence of one or more test smells. 
Test cases that have this pattern are identified just by re-executing  the code. This process is expensive and time-consuming, as demonstrated by \citet{Pinto-etal:2020} that found around 70\% of flaky test cases passed in more than 90\% of the executions. 
This fact serves as a motivation to use test smells as predictors of test flakiness. 

\section{Related Work}
\label{sec:relatedWork}

The presence of flaky tests may imply extra effort during the software engineering process. 
For instance, the developer can spend a lot of time debugging a failing test case that happens to be flaky. 
On the other side, ignoring a failing test by misclassifying it as flaky would cause the shipping of faulty software to the users.
Flaky tests may occur due to software changes, though other causes  have been identified in the literature~\cite{Luo-etal:2014, Eck-etal:2019}; see Table~\ref{tab:causes-flakiness}. 



In general, test flakiness detection has brought a lot of attention from industry and academia~\cite{Micco:2016, Fowler:2011, Palmer:2019, Goddard:2018,Zolfaghari-etal:2020}.
One direction is to adopt dynamic approaches whose core involves rerunning test cases for a fixed number of times~\cite{Lam-etal:2019, Shi-etal:2019}. 
A clear disadvantage is the cost of execution; for large test suites, this strategy may not scale. 
A different direction is to rely on statically-extracted information.


Works in the literature that are most related to ours apply Machine Learning (ML) methods to detect flaky tests~\cite{Alshammari-etal:2021,Atif-etal:2017,TariqKing-etal:2018,Pinto-etal:2020,Camara-etal:2021}. 
\citet{Atif-etal:2017} introduce an approach to minimize the workload of the test automation platform in Google.
To do so, the approach avoids the execution of test cases with low failure probability, and present insights to developers in order to prevent bugs. 
For ML part, the following features are used: CI tool, transitions PASSED-to-FAILED, fixes FAILED-to-PASSED, and developers' activity. 
The results show that the number of test runs can be reduced while maintaining similar bug detection capabilities. 

\citet{TariqKing-etal:2018} apply Bayesian networks to predict flaky tests. 
The used features are a mix of static and dynamic metrics: (i) complexity: assertion count, test class/method size, depth of inheritance tree, (ii) implementation coupling: coupling between objects and selector stability index, (iii) non-determinism: cyclomatic complexity and explicit wait count, (iv) performance: average execution time, and (v) general stability metrics: failure rate and flip rate. 
The authors evaluated the proposed approach with a case study with five teams developing a proprietary Web application. 
During the study, the approach supported the reduction of flaky tests; for some cases, the reduction was up to 60\%. 
Overall, the accuracy of the prediction was 65.7\%. 

Pinto et al.~\cite{Pinto-etal:2020} propose that there exists a vocabulary of patterns (words) in the test code that can be extracted using NLP to predict whether a test is flaky or not. 
To evaluate the approach, the authors constructed a dataset of Java projects, with test cases labelled as flaky and non-flaky, to train and test ML algorithms. Overall, all classifiers had good performance: Random Forest with the best precision (0.99) and F$_1$-Score, and SVM with the best recall (0.92).
The paper also shows the top-20 features with the highest information gain. 
Other studies also investigated how the vocabulary-based approach performs in Python projects~\cite{Ahmad2020,Haben2021}.

In a previous paper~\cite{Camara-etal:2021}, we conducted a replication of the \citet{Pinto-etal:2020}'s study.
The main extension was to assess the trained classifiers to predict flaky tests using a different test dataset (not used for training) in two contexts: For the intra-project context, the dataset contained tests from the same projects used during the training, while tests from different projects were used for the inter-project context.
Among the trained classifiers, the best one was LDA with recall of 0.75 for intra-project, and  0.45 for inter-project.
The authors conclude that the vocabulary-based approach is context sensitive and prone to overfitting.

In the same line, \citet{Alshammari-etal:2021} introduced an approach called FlakeFlagger that employs static and dynamic features like test smells, test coverage, source code management system, and source code, to predict test flakiness. 
Using a newly-defined dataset with 24 open source Java projects, the authors compared FlakeFlagger with the vocabulary-based approach~\cite{Pinto-etal:2020}, and a combination of both.
The obtained recall was similar for the 3 approaches compared (74\%, 72\%, and 74\%), but the precision has a 49\% difference between the FlakeFlagger and vocabulary-based, and the combined approach has a sensitive improvement of 6\%. 
So, the precision of the vocabulary-based approach was lower than FlakeFlagger. 
As for test smells, FlakeFlagger implements its own detector using an expanded definition of existing test smells. 
By analyzing the information gain, the results showed very low correlation with test flakiness.

We can see that the use of test smells is few explored in the literature.  We find only one work that consider test smells ~\cite{Alshammari-etal:2021}, but in combination with other predictors, such as coverage that need at least one execution of the code. 
Differently from related work,  we adopt only static metrics and perform a validation of the obtained models in  cross-project scenario.

\section{Methodology}
\label{sec:methodology}

The main goal of this study is to investigate the use of test smells as predictors of flakiness. The main advantage of this approach is that the smells can be collected statically. In addition to this, we investigate the use of the smell-based models as an alternative for a cross-project scenario.

\subsection {Research questions}
According to our goals we defined three Research Questions (RQs).

\begin{itemize}

    \item \textbf{RQ$_1$.} How accurately can we predict test flakiness based on test smells in the test cases? The goal of this question is to evaluate the performance of classifiers to predict test flakiness based on the presence/absence of test smells, without re-execution of the test suites.
   
    \item \textbf{RQ$_2$.} Which test smells are the most strongly associated with test flakiness prediction? The goal is to identify the test smells which are more related to flakiness in order to help development, code review, and debugging tasks.

    \item \textbf{RQ$_3$.} How does the test smell-based approach compare with the existing vocabulary-based approach? The goal was to compare the obtained results with the vocabulary-based approach~\cite{Pinto-etal:2020}, which can be considered as state-of-the art approach that also does not require test re-execution. 
    
\end{itemize}

RQ1 and RQ3 are answered considering two perspectives. In the first perspective a prediction model is built and evaluated according to some performance measures and compared with the vocabulary-based approach. For this end, the dataset made available by~\citet{Pinto-etal:2020} is used. In the second perspective, the evaluation considers the cross-project scenario. To this end, the dataset as extended by our previous work~\citet{Camara-etal:2021} is used. In this cross-project perspective, we evaluate the performance of the built models with different datasets: i) within a different set of test cases from the same software projects (intra-projects); and ii)  within other different projects (inter-projects).


The next sections present details about the methodology adopted to answer the RQs considering both perspectives: datasets, classifiers used to build the models, and evaluated measures.

\subsection{Datasets}
\label{subsec:datasets}

As mentioned before, for building the models, we use the dataset of Pinto et al.'s work~\cite{Pinto-etal:2020}, built based on 24 DeFlaker projects~\cite{Bell-etal:2018}.  The raw dataset has 49,919 test cases: 44,428 non-flaky, 5,069 flaky.  To evaluate the models in the cross-project perspective we used the data from a previous work~\cite{Camara-etal:2021}, built based on idFlakies projects of ~\citet{Lam-etal:2019}. This dataset contains only flaky tests, in a total of 422, from 72 different projects.

Both datasets were submitted to the tsDetect tool~\cite{Peruma-etal:2020}. In a first step this tool requires, for each test case, the identification of the corresponding production code, to then detect the smells the test contains. But for some test cases the code could not be identified, and as a consequence, the smell detection step could not be performed. Because of this, these test cases were removed from both datasets. At the end the dataset from Pinto et al.'s work~\cite{Pinto-etal:2020} dropped to 14,390 samples (11,319 non-flaky, 2,914 flaky), to deal with the imbalanced dataset, a number of non-flaky tests was selected randomly, equal to the number of flaky tests. In the dataset based on idFlakies projects~\citet{Lam-etal:2019}, the resultant sample dropped to 155.

As result we obtained a list with 19 test smells we used as features for the model; the test smells are described in \Cref{tab:test-smells}. The information about the smells is then augmented with two numerical features, acting as proxies of code complexity: LOC: number of lines of code of the test case; and smells count: total number of test smells present in the test case.  The vocabulary-based approach was applied according to related work~\cite{Camara-etal:2021}.

Following the methodology adopted in~\cite{Camara-etal:2021}, we generated  two datasets: one for training and testing  the models, and other to cross-project validation. The training and testing dataset is balanced and contains 2,777 samples, 1,377 flaky and 1,400 non-flaky, from 22 projects. This sample size was defined to be compatible with Pinto et al.'s work~\cite{Pinto-etal:2020} so we can use that work as benchmark.  The cross-project validation dataset was divided to attend the intra- and inter-project context. To obtain the intra-project dataset we filtered a set of flaky tests from 24 projects of the training dataset. At the end this set was composed by 35 samples. We obtained the dataset for the inter-project context by filtering out tests from projects present in the training dataset, this set are composed by 120 samples. In both  only flaky samples are present. 




\subsection{Used Classifiers}

We use eight classifiers available in the framework Scikit-learn~\cite{Pedregosa-etal:2011}: Random Forest, Decision Tree (DT), Naive Bayes, Support Vector Machine (SVM), Logistic Regression (LR), Linear Discriminant Analysis (LDA), K-Nearest Neighbour (KNN), and Perceptron. For comparison reasons, the choice of these classifiers, as well as the used parameters are based on related work~\cite{Pinto-etal:2020, Camara-etal:2021}.

\subsection{Evaluated metrics}

To evaluate the performance of the classifiers, the dataset was split into 80\% for training and 20\% for testing. 
We used the following standard metrics:
\begin{itemize}
    \item precision: the number of correctly classified flaky tests divided by the total number of tests that are classified as flaky;
    \item recall: the number of correctly classified flaky tests divided by the total number of actual flaky tests in the test set;
    \item F$_1$-Score: the harmonic mean of precision and recall;
    \item MCC (Matthews correlation coefficient): measures the correlation between predicted classes (i.e., flaky vs. non-flaky) and ground truth. Values of MCC vary in the interval of [-1,1], with 1 representing a perfect prediction; 
    \item AUC (area under the ROC curve):  measures the area under the curve which visualizes the trade-off between true-positive  and false-positive rates. 
\end{itemize}

Concerning the cross-project  perspective (namely, intra- and inter-project validation), the idFlakies dataset was adopted, and to evaluate the results only recall were used because, as mentioned, this dataset does not contain examples of non-flaky tests. To evaluate the relevance of the features (RQ2) we utilized the information gain (known as entropy), calculated for each output variable.  This value ranges from 0 (no gain) to 1 (maximum of information gain). It was calculated by using the method {\tt{mutual\_info\_classif}} of Scikit-learn with default settings.

\section{Analysis of Results}
\label{sec:results}

This section analyses the obtained results to answer the research questions.

\subsection{RQ$_{1}$ -- How accurately can we predict test flakiness based on test smells in the test cases?}

Following the experimental design described, we first built the prediction model by training and testing the classifiers. The results of the eight classifiers are presented in \Cref{tab:rq1}. All classifiers archived reasonably performance. Except by Naive-Bayes, they reached values greater than 70\% of precision, recall, F$_1$-Score, and AUC.  Again, except Naive-Bayes, all the classifiers obtained MCC values greater than 0.5 (which are close to 1 that represents a perfect classification).

The obtained results show that test smell-based models have reasonable performance to predict test flakiness, with precision values varying from 74 to 83\%. 
The best classifier was obtained with Random Forest and Decision Tree, which reached similar values for all measures,  with precision value of 83\%. Naive-Bayes presented the worst performance. 

\begin{table}[h]
\centering
\caption{Test smells-based classifiers' performance.}
\label{tab:rq1}
\begin{tabular}{|l|l|l|l|l|l|}

\hline
\textbf{Algorithm} & \textbf{Prec} & \textbf{Rec} & \textbf{F$_{1}$} & \textbf{MCC} & \textbf{AUC} \\ 

\hline
    Random Forest & \textbf{0.83} & \textbf{0.83} & \textbf{0.83} &         0.65  & \textbf{0.90} \\ \hline
    Decision Tree & \textbf{0.83} & \textbf{0.83} & \textbf{0.83} & \textbf{0.66} &         0.86  \\ \hline
    KNN           &         0.81  &         0.81  &         0.81  &         0.62  &         0.81  \\ \hline
    LR            &         0.79  &         0.79  &         0.79  &         0.59  &         0.87  \\ \hline
    LDA           &         0.78  &         0.78  &         0.78  &         0.56  &         0.86  \\ \hline
    Perceptron    &         0.78  &         0.78  &         0.78  &         0.55  &         0.86  \\  \hline
    SVM           &         0.75  &         0.75  &         0.75  &         0.50  &         0.83  \\ \hline
    Naive Bayes   &         0.74  &         0.65  &         0.61  &         0.37  &         0.78  \\ \hline
    
\end{tabular}
\end{table}

To validate the performance of the model in the cross-project context we tested the trained classifiers utilizing the flaky tests identified in the idFlakies dataset~\cite{Lam-etal:2019} (cross-validation dataset).
\Cref{tab:rq1_intra_inter} shows the results of intra- and inter-project contexts: it presents the classifiers' performance considering the recall, true positives (TP), and false negatives (FN).

\begin{table}[h]
\centering
\caption{Cross-project test smells-based classification.}
\label{tab:rq1_intra_inter}
\begin{tabular}{|l|l|l|l|l|l|l|}
\hline
\multirow{2}{*}{\textbf{Algorithm}} & \multicolumn{3}{c|}{\textbf{Intra-Project}} & \multicolumn{3}{c|}{\textbf{Inter-Project}} \\ \cline{2-7} 
                                    & \textbf{Rec}  & \textbf{TP}  & \textbf{FN}  & \textbf{Rec}  & \textbf{TP}  & \textbf{FN}  \\ \hline
Random Forest & 0.69          & 24 & 11 & 0.54          & 65 & 55  \\ \hline
Decision Tree & 0.66          & 23 & 12 & 0.54          & 65 & 55  \\ \hline
KNN           & 0.51          & 18 & 17 & 0.51          & 61 & 59  \\ \hline
LR            & \textbf{0.74} & 26 & 9  & 0.48          & 57 & 63  \\ \hline
LDA           & 0.66          & 23 & 12 & 0.48          & 57 & 63  \\ \hline
Perceptron    & 0.71          & 25 & 10 & 0.48          & 57 & 63  \\ \hline
SVM           & 0.66          & 23 & 12 & \textbf{0.55} & 66 & 54  \\ \hline
Naive Bayes   & 0.57          & 20 & 15 & 0.14          & 17 & 103 \\ \hline
\end{tabular}%
\end{table}

Considering the intra-project context, the performance of all classifiers dropped to recall values varying from 51\% to 74\%, being the best value reached by LR that classified correctly 26 of 9 flaky tests. The performance of the classifiers in the inter-project context dropped more significantly. But, excluding Naive-Bayes that reached the  value of 14\%, there is not great difference between the classifiers, with recall values varying from 48\% to 55\%. SVM reached the best performance.

\vspace{0.3cm}
\begin{tcolorbox}
\noindent {\bf Answer to RQ$_{1}$}: The obtained results show that test smells can be used as  predictors of flakiness. Nevertheless, the performance drops considerably in the inter-project context.
\end{tcolorbox}

\vspace{0.2cm}
\noindent {\bf Implications}: The results show that the smell-based models present performance comparable and in some cases better that some values reported in the literature (precision around 70\%)~\cite{TariqKing-etal:2018, Alshammari-etal:2021,Camara-etal:2021}. This lead to the conclusion that smells are good predictors of flakiness, but future work should explore the use of smells with other features, obtained statically. This can contribute  for improving performance in the cross-project validation  and for obtaining more generalizable models.
\vspace{0.2cm}

\subsection{RQ$_{2}$ -- Which test smells are the most strongly associated with test flakiness prediction?}

The  features most associated with test flakiness were determined by calculating the information gain based on the entropy of the features. \Cref{tab:rq2} shows all features adopted in the proposed model, ordered by their relevance (i.e., information gain); the data is based on the training dataset. Column `Total' shows how many tests are affected by the feature, while columns `Flaky tests' and `Non-Flaky tests' bring the number of affected tests that are flaky or not, respectively; the percentages are presented in the next columns.

\begin{table*}[h]
\centering
\caption{Information gain of each feature of the model.}
\label{tab:rq2}

\begin{tabular}{|l|l|l|l|l|l|l|l|}

\hline

\textbf{Pos.} &
  \textbf{Features} &
  \textbf{\begin{tabular}[c]{@{}l@{}}Inf.   \\      Gain\end{tabular}} &
  \textbf{Total} & 
  \textbf{\begin{tabular}[c]{@{}l@{}}Flaky   \\      tests\end{tabular}} &
  \textbf{\begin{tabular}[c]{@{}l@{}}\% Flaky   \\      tests\end{tabular}} &
  \textbf{\begin{tabular}[c]{@{}l@{}}Non-Flaky   \\      tests\end{tabular}} &
  \textbf{\begin{tabular}[c]{@{}l@{}}\% Non-Flaky   \\      tests\end{tabular}}  \\ \hline

1  & LOC                        & 0.2545 & 2777 & 1377 & 49.59\%  & 1400 & 50.41\% \\ \hline
2  & Assertion Roulette         & 0.0832 & 1389 & 968  & 69.69\%  & 421  & 30.31\% \\ \hline
3  & Smells Count               & 0.0271 & 2655 & 1356 & 51.07\%  & 1299 & 48.93\% \\ \hline
4  & Sleepy Test                & 0.0195 & 112  & 105  & 93.75\%  & 7    & 6.25\%  \\ \hline
5  & General Fixture            & 0.0160 & 267  & 61   & 22.85\%  & 206  & 77.15\% \\ \hline
6  & Duplicate Assert           & 0.0155 & 376  & 269  & 71.54\%  & 107  & 28.46\% \\ \hline
7  & Constructor Initialization & 0.0109 & 68   & 63   & 92.65\%  & 5    & 7.35\%  \\ \hline
8  & Redundant Print            & 0.0106 & 58   & 55   & 94.83\%  & 3    & 5.17\%  \\ \hline
9  & Sensitive Equality         & 0.0059 & 129  & 95   & 73.64\%  & 34   & 26.36\% \\ \hline
10 & Lazy Test                  & 0.0055 & 1788 & 817  & 45.69\%  & 971  & 54.31\% \\ \hline
11 & Resource Optimism          & 0.0043 & 75   & 17   & 22.67\%  & 58   & 77.33\% \\ \hline
12 & Conditional Test Logic     & 0.0042 & 356  & 219  & 61.52\%  & 137  & 38.48\% \\ \hline
13 & Unknown Test               & 0.0021 & 544  & 234  & 43.01\%  & 310  & 56.99\% \\ \hline
14 & Verbose Test               & 0.0018 & 7    & 7    & 100\%    & 0    & 0.00\%  \\ \hline
15 & Magic Number Test          & 0.0011 & 411  & 227  & 55.23\%  & 184  & 44.77\% \\ \hline
16 & Mystery Guest              & 0.0006 & 124  & 71   & 57.26\%  & 53   & 42.74\% \\ \hline
17 & Eager Test                 & 0.0003 & 970  & 496  & 51.13\%  & 474  & 48.87\% \\ \hline
18 & Redundant Assertion        & 0.0000 & 8    & 4    & 50.00\%  & 4    & 50.00\% \\ \hline
19 & Default Test               & 0.0000 & 0    & 0    & 0.00\%   & 0    & 0.00\%  \\ \hline
20 & Empty Test                 & 0.0000 & 0    & 0    & 0.00\%   & 0    & 0.00\%  \\ \hline
21 & Ignored Test               & 0.0000 & 0    & 0    & 0.00\%   & 0    & 0.00\%  \\ \hline

\end{tabular}%
\end{table*}

Considering the contribution of the test smells for the model, we observe four smells with at least 90\% of affected tests being flaky. They are: \texttt{Verbose Test} (100\%),  \texttt{Redundant Print} (94.83\%),  \texttt{Sleepy Test} (93.75\%), and  \texttt{Constructor Initialization} (92.65\%). Of these,  \texttt{Sleepy Test}, and  \texttt{Constructor Initialization}. These smells can be associated with types of flaky tests. 

\begin{sloppypar}
The \texttt{Sleepy Test} smell is related to the use of delays (\textsc{Thread.sleep()} or similar statements) to wait for other components to be ready to be executed. That behavior is described by \citet{Luo-etal:2014} as the flaky test type  \texttt{Async Wait}. In \Cref{tab:rq3_intra_inter} \texttt{Sleepy Test}  is found in 105 (94\%) flaky tests and in just 7 non-flaky tests.  \texttt{Constructor Initialization} occurs when a test class has a constructor, which also can be related to the flakiness type  \texttt{Test Order Dependency}, when the test result depends on the order in which the tests are run. This test smell type are found in 93\% of flaky tests and 6\% in non-flaky tests. 
The \texttt{Assertion Roulette} smell occurs when the method has more than an assertion; this is the test smell most found in our study, in 69.69\% of the cases  they are associated with flaky tests and is the second feature with biggest information gain. The average of lines of code of methods that are identified with \text{Assertion Roulette} is over 28, containing 3.4 smells. 
\end{sloppypar}

\Cref{tab:smells-count} shows the smell count distribution over the training dataset, in which we can see the greater the number of smells the great the  flakiness percentage. In our dataset we have 122 test cases without test smells, 17\% are identified as flaky, when considering test cases with one smell, the percentage grows to 39\%, becoming at 71\% with six smells. However, if a group with 1 to 4 test smells is considered we have a set of 89\% of the data, 50\% of flaky, and 50\% of non-flaky tests, that is, future work is necessary to better investigate if  the feature smell count is a good feature to identify flaky test.  

\begin{table}[h]
\centering
\caption{Smell count distribution.}
\label{tab:smells-count}
\begin{tabular}{|l|l|l|l|l|}
\hline
\textbf{\begin{tabular}[c]{@{}l@{}}Smells\\ Count\end{tabular}} &
  \textbf{Non-Flaky} &
  \textbf{\begin{tabular}[c]{@{}l@{}}\% \\ Non-Flaky\end{tabular}} &
  \textbf{Flaky} &
  \textbf{\begin{tabular}[c]{@{}l@{}}\% \\ Flaky\end{tabular}} \\ \hline
0 & 101 & \textbf{83\%} & 21  & 17\%          \\ \hline
1 & 469 & 61\%          & 305 & 39\%          \\ \hline
2 & 327 & 50\%          & 329 & 50\%          \\ \hline
3 & 264 & 46\%          & 308 & 4\%           \\ \hline
4 & 159 & 35\%          & 301 & 65\%          \\ \hline
5 & 63  & 46\%          & 73  & 54\%          \\ \hline
6 & 12  & 29\%          & 30  & \textbf{71\%} \\ \hline
7 & 4   & 33\%          & 8   & 67\%          \\ \hline
8 & 1   & 33\%          & 2   & 67\%          \\ \hline
\end{tabular}%
\end{table}

\vspace{0.3cm}
\begin{tcolorbox}
\noindent {\bf Answer to RQ$_{2}$}: \texttt{Sleepy Test}, and  \texttt{Constructor Initialization} can be associated with flaky test types. This is demonstrated in our dataset by the distribution of numbers in flaky tests and by the information gain. 
\end{tcolorbox}

\vspace{0.2cm}
\noindent {\bf Implications}: The information gain presents the ordered set of features that are more discriminant; it is possible to see some features that make sense but more studies are necessary to analyze if in fact all of them have a positive impact in the prediction.

\subsection{RQ$_{3}$ -- How does test the smell-based approach compare with the existing vocabulary-based approach?}

Using the vocabulary-based approach we trained the eight classifiers with our training and testing dataset (the same used in  RQ$_{1}$) to answer RQ$_{3}$. \Cref{tab:rq3} shows the results of the trained classifiers. Despite we are using distinct datasets, as explained in Section~\ref{subsec:datasets} the results are similar to those presented in previous studies~\cite{Camara-etal:2021,Pinto-etal:2020}. 

After this, we apply the  vocabulary-based approach using our cross-validation dataset. The results of the classifiers  obtained for the cross-project contexts are presented in \Cref{tab:rq3_intra_inter}. The obtained performance of intra-project classifications is worst than the performance reported previously~\cite{Camara-etal:2021} where the best classifier was LDA with 75\% of recall classifying 20 of out 60 samples correctly. Here, the best classifier was KNN with 57\% of recall, classifying correctly 20 of out 45 samples. The inter-project performance was slightly better than in~\cite{Camara-etal:2021}, obtaining 56\% of recall against 48\% (122 of 134).

\begin{table}[h]
\centering
\caption{Vocabulary-based classifiers' performance.}
\label{tab:rq3}
\begin{tabular}{|l|l|l|l|l|l|}

\hline
\textbf{Algorithm} & \textbf{Prec} & \textbf{Rec} & \textbf{F$_{1}$} & \textbf{MCC} & \textbf{AUC} \\ \hline

    Random Forest & \textbf{0.97} & \textbf{0.97}            & \textbf{0.97} &         0.93  & \textbf{0.99} \\ \hline
    Perceptron    & \textbf{0.97} & \textbf{0.97}            & \textbf{0.97} &         0.93  & \textbf{0.99} \\ \hline
    Decision Tree &         0.94  &         0.94             &         0.94  &         0.87  &         0.94  \\ \hline
    Naive Bayes   &         0.95  &         0.95             &         0.95  &         0.90  &         0.95  \\ \hline
    SVM           & \textbf{0.97} & \textbf{0.97}            & \textbf{0.97} & \textbf{0.94} & \textbf{0.99} \\ \hline
    LR            & \textbf{0.97} & \textbf{0.97}            & \textbf{0.97} & \textbf{0.94} & \textbf{0.99} \\ \hline
    LDA           &         0.87  &         0.87             &         0.87  &         0.74  &         0.88  \\ \hline
    KNN           &         0.93  &         0.93             &         0.93  &         0.86  &         0.93  \\ \hline

\end{tabular}
\end{table}

\begin{table}[h]
\centering
\caption{Cross-project vocabulary-based performance.}
\label{tab:rq3_intra_inter}
\begin{tabular}{|l|l|l|l|l|l|l|}
\hline
\multirow{2}{*}{\textbf{Algorithm}} & \multicolumn{3}{c|}{\textbf{Intra-Project}} & \multicolumn{3}{c|}{\textbf{Inter-Project}} \\ \cline{2-7} 
                                    & \textbf{Rec} & \textbf{TP} & \textbf{FN} & \textbf{Rec} & \textbf{TP} & \textbf{FN} \\ \hline
Decision Tree & 0.31          & 11          & 24 & 0.39          & 47          & 73  \\ \hline
LDA           & 0.29          & 10          & 25 & \textbf{0.56} & \textbf{67} & 53  \\ \hline
LR            & 0.20          & 7           & 28 & 0.30          & 36          & 84  \\ \hline
Random Forest & 0.17          & 6           & 29 & 0.29          & 35          & 85  \\ \hline
Naive Bayes   & 0.17          & 6           & 29 & 0.13          & 15          & 105 \\ \hline
SVM           & 0.09          & 3           & 32 & 0.17          & 20          & 100 \\ \hline
KNN           & \textbf{0.57} & \textbf{20} & 15 & 0.23          & 27          & 93  \\ \hline
Perceptron    & 0.34          & 12          & 23 & 0.33          & 40          & 80  \\ \hline
\end{tabular}%
\end{table}

Comparing with the smell-based approach, the vocabulary-based approach presents better performance: the best F$_{1}$ score of vo\-ca\-bu\-lary-based models is 97\% (Random Forest), while smell-based approach  has a score of 83\% (Random Forest). Analysing  MCC, the difference is higher. This measure takes into account true and false positives, and negatives, the best result obtained by the smell-based approach is 0.66 and the best obtained by the vocabulary-based approach is 0.94. 

However, the results obtained in the cross-project validation show that the test smell-based approach reaches better results. In the intra-project context, the test smell-based approach obtained 74\% of recall (LR), and the vocabulary-based one achieved at most 57\% (KNN).  Considering the best classifiers in the inter-project context, the performance is  similar, the smell-based approach obtained a recall of at most 55\% (SVM), against 56\% of the vocabulary-based (LDA). But if we consider all the classifiers, we observe that  smell-based approach has a general better performance in the cross-project validation than the vocabulary-based approach. 

\vspace{0.3cm}

\begin{tcolorbox}
\noindent {\bf Answer to RQ$_{3}$}: The performance of the vocabulary-based models are better than the performance of the test smell-based ones using the training and testing dataset.  But in the cross-project validation scenario, the  smell-based approach obtains significant better results in the intra-project and inter-project contexts. 
\end{tcolorbox}

\vspace{0.2cm}
\noindent {\bf Implications}: The results obtained in our study corroborate the ones presented in our previous work~\cite{Camara-etal:2021}. The  vocabulary-based approach do not have a good performance in cross-project validation.  Our results shows that the use of test smells can be a good alternative for overcoming some limitations of the vocabulary-based approach, and  smell-based models can be more generalizable.  
\vspace{0.2cm}

\section{Threats to Validity}
\label{sec:threats}

Threats to construct validity are related to the metrics used to evaluate the results. But to minimize this threat we adopted the most common  measures used in the ML field to evaluate the classifiers.

During the test code pre-processing to obtain the the test smells in some cases tsDetect~\cite{Peruma-etal:2020} could not identify the production class, then the smells could not be extracted, which can compromise the result. 

Threats to internal validity may comprise the results when relating independent and dependent variables. The absence of non-flaky class on the cross-project dataset made impossible to obtain precision and other metrics to use as a benchmark. This should be evaluated  in future work.  

External validity is connected to the generalization of the obtained results. 
As in similar studies, we cannot generalize the results, once this study targeted the Java language and a limited set of project domains. 
Considering the cross-project validation, the results obtained in the intra-project context show an expressive reduction in the performance if compared with the results reported in~\cite{Camara-etal:2021}. This could be a result of the  cross-project validation dataset, which is smaller and should be increased in future works to better understand the performance of prediction.
\section{Conclusion}
\label{sec:conclusion}

Regression tests are an important practice for supporting continuous integration and delivery,  and flaky tests are disturbing to this process. 
Consequently, the proper identification and prevention of test flakiness are important topics pursued by researchers and practitioners. 

This paper investigated the use of test smells as features to predict flaky tests. To this end, we conducted an empirical study to evaluate the performance of the test smell-based approach for  prediction of flakiness. Such a performance was measured considering different evaluation indicators, and  cross-project validation. We also identified the test smells most strongly associated with test flakiness prediction, and compared the test smell-based approach with the state-of-the art  vocabulary-based approach. 



As result, we observed that test smells are a potentially good predictors of test flakiness; the results of intra- and inter-project are promising. 
Some test smells like {\it Sleepy Test} and {\it Constructor Initialization} are strongly associated with flakiness. Compared to the vocabulary-based approach, the test smell-based one obtained models that have in the best case, a precision 14\% lower. In the cross-project validation, the test smell-based approach performs better in the intra- and inter- project contexts. 



This study opens the possibility to use smell-based models for the prediction of test flakiness. As a future work it is possible to expand the training dataset by adding samples of other projects of other contexts and programming languages. We also intend to explore other static and dynamic features in combination with test smells.

\section*{Acknowledgment}
This work is partially supported by CNPq (Andre T. Endo grant nr. 420363/2018-1 and Silvia Regina Vergilio grant nr. 305968/2018-1) and by IN2 Institute (Bruno Henrique Pachulski Camara grant nr. 002/2021).

\balance
\bibliographystyle{IEEEtranN}
\bibliography{bib}

\end{document}